\documentclass[sigplan]{acmart}

\def \arxiv {arxiv}

\usepackage{balance}
\usepackage{booktabs}
\usepackage{enumitem}
\usepackage{xspace}
\usepackage{listings}
\usepackage[symbol]{footmisc}
\setcopyright{rightsretained} \acmConference[]{}{}{}
\acmYear{2018}

\newcommand{\treg}{\texttt{r15}\xspace}

\begin{document}
\title{Mitigating Branch-Shadowing Attacks on Intel SGX \\using Control Flow Randomization}  
\author{Shohreh Hosseinzadeh}
\authornote{\label{authorThing}S. Hosseinzadeh and H. Liljestrand contributed equally to this work, which was done while S. Hosseinzadeh was visiting the Secure Systems Group at Aalto University.}
\orcid{1234-5678-9012}
\affiliation{\institution{University of Turku, Finland}}
\email{shohos@utu.fi}

\author{Hans Liljestrand\textsuperscript{\ref{authorThing}}}
\affiliation{  \institution{Aalto University, Finland}
        }
\email{hans.liljestrand@aalto.fi}

\author{Ville Lepp{\"a}nen}
\orcid{1234-5678-9012}
\affiliation{  \institution{University of Turku, Finland}
        }
\email{ville.leppanen@utu.fi}

\author{Andrew Paverd}
\affiliation{  \institution{Aalto University, Finland}
        }
\email{andrew.paverd@ieee.org}

\renewcommand{\shortauthors}{S. Hosseinzadeh et al.}

\begin{abstract}
Intel Software Guard Extensions (SGX) is a promising hardware-based technology for protecting sensitive computations from potentially compromised system software.
However, recent research has shown that SGX is vulnerable to \emph{branch-shadowing} -- a side channel attack that leaks the fine-grained (branch granularity) control flow of an \emph{enclave} (SGX protected code), potentially revealing sensitive data to the attacker.
The previously-proposed defense mechanism, called \emph{Zigzagger}, attempted to hide the control flow, but has been shown to be ineffective if the attacker can single-step through the enclave using the recent SGX-Step framework. 

Taking into account these stronger attacker capabilities, we propose a new defense against branch-shadowing, based on control flow randomization.
Our scheme is inspired by Zigzagger, but provides quantifiable security guarantees with respect to a tunable security parameter. 
Specifically, we eliminate conditional branches and hide the targets of unconditional branches using a combination of compile-time modifications and run-time code randomization.

We evaluated the performance of our approach by measuring the run-time overhead of ten benchmark programs of SGX-Nbench in SGX environment.

\end{abstract}



\maketitle

\section{Introduction}

Intel Software Guard Extension (SGX)~\cite{Intel-SGX} is a recent hardware-based Trusted Execution Environment (TEE) providing secure computation and guaranteeing the integrity and confidentiality of data within an \emph{enclave}.
The enclave is protected from all other software on the platform, including potentially malicious system software (e.g., operating system, hypervisor, and BIOS). 
Additionally, SGX enables hardware-based measurement, attestation, and integrity verification of application code. 

Although Intel has stated that side-channel attacks are beyond the scope of SGX~\cite{Intel-side-channels}, recent research has demonstrated that SGX is susceptible to several side-channel attacks, which could leak secret information. They have also proposed defenses for making SGX enclave resistant to side channels.
In particular, Lee et al.~\cite{SGX-Branch-Shadowing} demonstrated a \emph{branch-shadowing} side channel attack that allows untrusted software to learn the precise control flow of code running inside an enclave.
If this control flow depends on any secret information, this side channel would leak the secret information.
Fundamentally, the attack works by abusing the CPU's Branch Prediction Unit (BPU).
The BPU is used to improve performance by allowing pipelining of instructions before exact branching decisions are known, i.e., whether branches are taken or not and the target of indirect branches.
The BPU bases its decisions on recent branch history, which is stored in the CPU's internal Branch Target Buffer (BTB).
Two critical factors allow this attack to proceed: 1) BTB entries created by branches inside the enclave are not cleared when the enclave exits; and 2) BTB entries only contain the lower 31 bits of the branch instruction's address, allowing the attacker to create \emph{shadow} branch instructions outside the enclave that map to the same BTB entries as the enclave's branches.
The attacker executes the victim enclave, interrupts it immediately after the branch instruction, executes his shadow branch code and checks whether the branches were correctly predicted, thus revealing whether the BTB entry had been created by the enclave.

Lee et al.~\cite{SGX-Branch-Shadowing} also proposed a software-based defense against the branch-shadowing attack, called \emph{Zigzagger}.
Using compile-time instrumentation, Zigzagger converts all conditional and unconditional branches into unconditional branches targeting Zigzagger's trampolines, i.e., minimal code sections that hold intermediate jumps --- bounces --- to the target locations. 
The Zigzagger trampolines initiate a series of jumps back-and-forth to different branches.
The idea is that the attacker cannot interrupt the enclave with sufficient precision to shadow the target branch in this rapid series of jumps.
However, the recent SGX-Step framework~\cite{SGX-Step} invalidates this assumption by showing how an enclave can be interrupted with single instruction granularity, thus breaking the Zigzagger defense.

The recent Spectre~\cite{Spectre} attacks, and their subsequent SGX-specific SGXPectre variant~\cite{SgxPectre} are similar to branch-shadowing in that they exploit the BPU.
However, we have confirmed experimentally that neither recent firmware patches~\cite{Intel-mitigations}, nor the Retpoline compiler-based mitigation~\cite{Intel2018retpoline} affect the ability to perform branch-shadowing attacks.

To overcome this challenge, we present a new defense against branch-shadowing, even if the attacker can single-step through the enclave.
Similar to Zigzagger, we use compile-time modifications to convert all branch instructions into unconditional branches targeting our in-enclave trampoline code.
At run-time, we then randomize the layout of our trampoline, forcing the attacker to shadow all possible locations.
The finite size of the BTB limits the number of guesses the attacker can perform, and thus we can quantify and limit the success probability of a branch-showing attack using the size of the trampoline as a tunable security parameter.

Our contributions are therefore:

\begin{itemize}[wide=0pt, leftmargin=\dimexpr\labelwidth + 2\labelsep\relax]
    \item Experimental analysis demonstrating that the recent Spectre mitigation techniques \emph{do not} affect the branch-shadowing attack (Section~\ref{sec:Spectre-mitigation}).
    \item A new approach for defending against branch-shadowing attacks, even in the presence of single-step enclave execution, using control flow randomization (Section~\ref{sec:proposed-scheme}).
      \item An initial LLVM implementation of our solution (Section~\ref{sec:imp-details})
    \ifdefined\arxiv
    \else
    \footnote{Available online at \url{https://bit.ly/2zwS1p8}}
    \fi
    and a quantitative evaluation of its performance and security guarantees (Section~\ref{sec:evaluation}).
\end{itemize}

\section{Background}
\label{sec:background}

\subsection{Branch Prediction}

Intel CPUs use instruction pipelining to load and execute instructions in batches. 
This allows optimization such as parallelizing and reordering of instructions.
The CPU also performs speculative execution, i.e., it uses the BPU to predict which branches will be taken, and executes them before knowing if they are taken~\cite{Intel-optimization-manual}.
In modern microprocessors, the BPU typically consists of two main subsystems, a BTB and a directional predictor. 

The BTB is used to predict the targets of indirect branches~\cite{intelManual}.
Whenever a branch is taken, a new record is created in the BTB associating the branch instruction's addresses with the target address.
Upon encountering subsequent branch instructions, the BPU checks the BTB for the branch instruction address and, if an entry exists, it predicts that the current branch instruction will behave in the same way.
The exact details of the BTB lookup algorithms, hashing and size are not public, but the BTB size on Intel Skylake CPUs has been experimentally determined to be 4096 entries~\cite{SGX-Branch-Shadowing}. 
The directional predictor is used to predict whether or not a conditional branch will be taken.
It stores the pattern of the previously taken branches in a Pattern History Table (PHT), based on which it predicts whether or not the current branch will be taken~\cite{BranchScope}.

Multiple processes executing on the same core share the same BPU. 
On one hand, this is favorable from a complexity and utilization point of view, but on the other hand, it allows an attacker to misuse the BPU across cores and infer the target and direction of branch instructions~\cite{SGX-Branch-Shadowing, BranchScope}.

\subsection{Intel SGX}

Intel SGX is an instruction set extension that provides new instructions to instantiate Trusted Execution Environments (TEEs), called \emph{enclaves}, consisting of code and data.
An enclave's data can only be accessed by code running within the enclave, thus protecting it from all other software on the platform, including privileged system software such as the OS or hypervisor.
Enclave data is automatically encrypted before it leaves the CPU boundary.
However, the OS remains in control of process scheduling and memory mapping, and can therefore control the mapping of (encrypted) enclave memory pages and interrupt enclave execution.

\subsection{Branch-shadowing Attacks on SGX}

Lee et al.~\cite{SGX-Branch-Shadowing} presented a \emph{branch-shadowing} attack against Intel SGX.
It is a type of side-channel attack that leaks the fine-grained control flow (i.e., at branch level) of code running inside an enclave.
The attack can be used to infer branching decisions, i.e., whether a (conditional/unconditional/indirect) branch has been taken or not.
If this control flow depends on confidential information held by the enclave, the attacker would be able to infer this information from the leaked control flow.
Because SGX does not clear the BPU on enclave exit an attacker outside the enclave could probe the BPU to infer control flow decisions taken within.
 The attacker first statically analyzes the unencrypted enclave code and enumerates all branches (i.e., conditional, unconditional, and indirect) together with their target addresses. 
She then creates shadow code where the branch-instructions and target addresses are aligned such that they will use the same BPU history entries.
The attacker then allows the enclave to execute briefly before interrupting it.
Finally, she enables the performance counter, in particular the Last Branch Record (LBR), and executes the shadow code, prompting the CPU to predict shadow-branch behavior based on prior enclave execution.
The LBR contains information on branch prediction but cannot record in-enclave branches.
However, the in-enclave branches can be inferred from the LBR entries for the branches executed after exiting the enclave.
Unlike cache-based channels, this does not require timing because the LBR directly reports prediction status.

\subsection{Zigzagger and SGX-Step}

Lee et al.~\cite{SGX-Branch-Shadowing} also presented \textit{Zigzagger}, a software-based countermeasure to thwart the attack. 
Zigzagger removes the branches from the enclave functions by obfuscating and replacing a set of branch instructions with a series of indirect jumps. 
Instead of each conditional branching instruction, an indirect jump and a conditional move (CMOV) is used.
Zigzagger assumes that an attacker cannot precisely time the enclave interrupts, i.e., a single probe will cover over 50 instructions. 
It introduces a trampoline to exercise all unconditional jumps before finally jumping to the final destination. 
The attacker will typically always detect the same set of taken jumps (i.e., all the unconditional jumps) and cannot distinguish the final jump from the decoy-jumps.

However, Van Bulck et al.~\cite{SGX-Step} presented \textit{SGX-Step}, a framework consisting of a Linux kernel driver and runtime library that manipulates the processor's Advanced Programmable Interrupt Controller (APIC) timer in order to interrupt an enclave after a single instruction i.e., to single-step the enclave's execution.
They show that this makes the Zigzagger defense ineffective because the attacker can distinguish meaningful jumps from decoys.

\section{Spectre Mitigation Techniques}
\label{sec:Spectre-mitigation}

The recent Spectre~\cite{Spectre} and SGXPectre~\cite{SgxPectre} attacks are similar to branch-shadowing in that they abuse the BPU to exploit speculative execution.
But whereas branch-shadowing aims to infer prior branching behavior, these attacks instead manipulate upcoming branch prediction, e.g., cause speculative execution to touch otherwise inaccessible memory.
Although not designed to do so, we suspected the new Spectre mitigation techniques could also affect the branch-shadowing attacks.
However, our testing indicates that neither the recent firmware patches from Intel~\cite{Intel-mitigations}, nor the compiler-based Retpoline~\cite{Intel2018retpoline} affect the ability to perform branch-shadowing attacks against SGX. 

In particular, we confirmed that Indirect Branch Restricted Speculation (IBRS) --- designed to prevent unprivileged code from affecting speculation in privileged execution, e.g., within the enclave --- has no effect on branch-shadowing.
In our tests we saw no difference between an updated i7-7500U CPU and non-updated machines.
We speculate that this is because IBRS is specifically designed low-privilege code from affecting high-privilege code, not the branch-shadowing scenario.
The Retpoline defense replaces branch instructions with return instructions but our tests indicate that return statements affect the BTB, not only the dedicated Return Stack Buffer (RSB). 
SGXPectre further demonstrated that Spectre attacks can be performed against Retpoline.

\section{Attacker Model and Requirements}
\label{sec:threat-model}

We assume that the attacker has fine-grained control of enclave execution, i.e., can interrupt the enclave with instruction-level accuracy.
The attacker can thus perform a branch-shadowing attack against every branch instruction.
Specifically, the attacker can determine whether or not a branch instruction has been executed and taken (i.e., whether a conditional jump fell through or not).
If the branching decisions depend on sensitive enclave data, the attacker can infer this data through the branch-shadowing attack.

This is a significantly stronger attacker capability than that assumed by previous work~\cite{SGX-Branch-Shadowing} because Van Bulck et al.~\cite{SGX-Step} showed that single-step execution of SGX enclaves is both feasible to implement and sufficient to break existing defenses like Zigzagger~\cite{SGX-Branch-Shadowing}.
We focus on branch-shadowing attacks and do not consider other side-channels, such as cache or page-fault attacks.

Given these attacker capabilities, we require a defence mechanism that prevents fine-grained branch-shadowing from revealing secret-dependent control flow.
Specifically, we require that branch instructions in the instrumented code are:

\begin{enumerate}[wide=0pt, leftmargin=\dimexpr\labelwidth + 4\labelsep\relax,label=\textbf{R.\arabic*}]
    \item \label{reqObservable}     Any branch that can be directly observed through branch-shadowing reveals no secret-dependent control flow information.
    \item \label{reqNonObservable}     For any secret-dependent branches, the attacker's probability of success is bounded based on a security parameter $\mathtt{k}$.
\end{enumerate}

\section{Proposed Approach}
\label{sec:proposed-scheme}

Our mitigation scheme uses compile-time obfuscation and run-time randomization to hide the control flow of an enclave application. 
While our proposed method is inspired by and uses a similar approach to Zigzagger, we assume a stronger attacker model.
Specifically, our approach can defend against branch-shadowing even in the presence of an attacker with single-step capabilities.

\begin{figure} [!h]
\centering
\includegraphics[width=0.9\columnwidth]{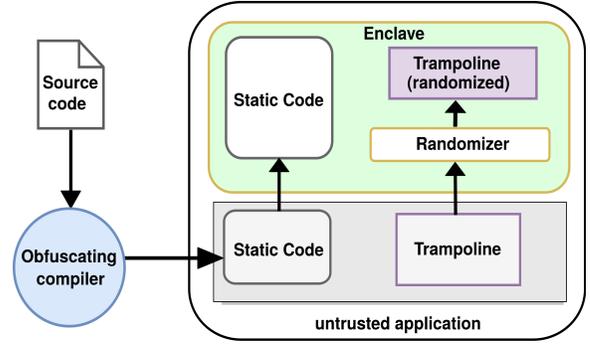}
\caption{System design}
\label{system-design}
\end{figure}

Figure \ref{system-design} illustrates the high-level view of our approach. 
The system consists of two main components: an obfuscating compiler and a run-time randomization library. 
The compiler generates the obfuscated code and trampolines, and the randomization library randomizes the trampolines at each enclave entry. 
This randomization only affects the trampolines allowing most code to remain securely in execute-only memory.

The obfuscating compiler modifies the code by converting all branching instructions to indirect branches.
The indirect branch targets are then explicitly set by the instrumentation depending on the converted branch type.
We use conditional moves as replacements for conditional branches, allowing us to replicate the functionality of any conditional branch without involving the BPU\@. The observable control flow transitions, i.e., non trampoline branches, are further organized so that they are always unconditionally executed in the same order.

The key insight of our approach is that, unlike Zigzagger, the trampolines are randomized inside the enclave at run-time. 
This prevents the attacker from reliably tracking their execution. 
Taken together, these two properties fulfill requirements \ref{reqObservable} and \ref{reqNonObservable}, as we show in our security evaluation in Section~\ref{sec:evaluation}.

\lstset{language=C}
\lstset{escapeinside={(*@}{@*)}}
\lstset{frame = single, xleftmargin=5mm, framexleftmargin=4mm}
\begin{lstlisting}[float,floatplacement=t,label={lst:exampleCode},caption={Example code before instrumentation}]
if (a !=0) {
    /* (*@\texttt{Block1}@*) */
} else {
    /* (*@\texttt{Block2}@*) */
}
/* (*@\texttt{Block3}@*) */
\end{lstlisting}

\begin{figure}[h]
\centering
\includegraphics[width=0.65\columnwidth]{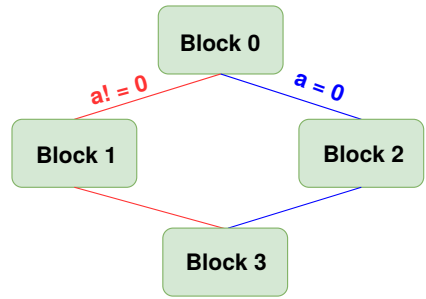}
\caption{Original control flow graph}
\label{original-graph}
\end{figure}

\begin{figure}[!h]
\centering
\includegraphics[width=0.65\columnwidth]{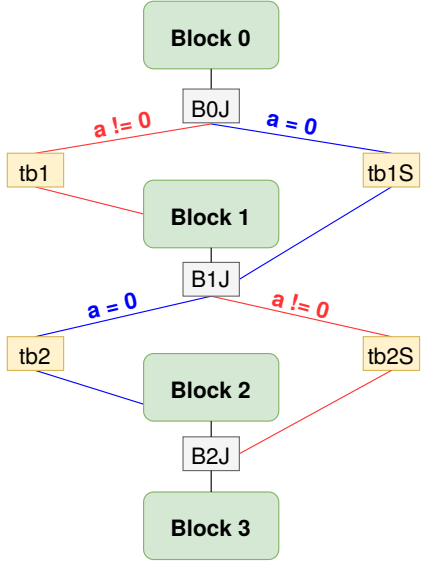}
\caption{Modified control flow graph}
\label{modified-graph}
\end{figure}

\begin{figure} [!h]
\centering
\includegraphics[width=0.9\columnwidth]{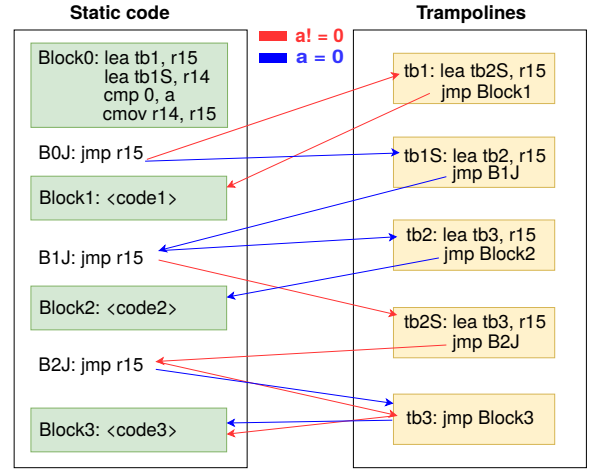}
\caption{Modified code protected by our approach}
\label{trampoline}
\end{figure}

Listing~\ref{lst:exampleCode} and Figure~\ref{original-graph} show a single \texttt{if}-statement and corresponding Control Flow Graph. The corresponding obfuscated CFG is show in Figure~\ref{modified-graph}.
Figure~\ref{trampoline} shows the same obfuscated code with the branch instruction converted. 
The static code is produced at compile time and its layout is assumed to be known to the attacker. 
The trampoline is similarly produced at compile time but is then randomized at run-time within the enclave. 
We assume that the attacker can observe and shadow the static code whereas the trampoline is unknown.

Specifically, our approach works as follows:

\noindent\textbf{1)}
All branching instructions are converted to indirect unconditional branches.
A register (\treg) is reserved and populated with the original branch targets.
 The target addresses are taken from a jump-table that is updated during randomization. Conditional branches are converted to conditional moves (\texttt{cmov}) (e.g., \texttt{Block0} in Figure~\ref{trampoline}).

\noindent\textbf{2)}
Each block is followed by a \emph{jump-block} that jumps to a trampoline indicated by \treg.
Execution flows that do not include a specific block still go through any intermediate jump-blocks to ensure that all indirect jumps outside the trampolines are executed.
For instance, when taking the \texttt{if}-clause (\texttt{Block1}), the else-block (\texttt{Block2}) must not be executed but the corresponding jump-block (\texttt{B2J}) must be (e.g., the blue line in Figure~\ref{trampoline}).
This ensures that an attacker always sees the same sequence of jumps jumps (i.e., \texttt{B0J}, \texttt{B1J}, and \texttt{B2J}), regardless of actual executed code.

\noindent\textbf{3)}
The corresponding trampolines are created, corresponding to either the branching target or the fall-through block (i.e., the next block that will be executed when a conditional branch is not taken).
In Figure~\ref{modified-graph}, after execution of the if-block (\texttt{Block1}) the control flow is transferred to \texttt{tb2S} that will jump to the following jump-block \texttt{B2J} without executing the corresponding \texttt{Block2} itself.
  
\noindent\textbf{4)}
When skipping a block --- e.g., the \texttt{else} block after taking the \texttt{if} block --- we must nonetheless execute the corresponding jump-block to prevent its omission from leaking information.
The jump-block target is prepared in the prior trampoline block by setting \treg.
For instance, after executing the if-block the corresponding trampoline (\texttt{tb2S}) not only jumps to the correct jump-block, but also sets the next target, \texttt{tb3}, into \treg.
To prevent timing attacks that measure the number of instructions between jump-blocks the skipping trampolines are populated with dummy-instructions, which ensure that the timing between each jump-block is constant regardless of control flow.
Nested blocks, although not shown in our example code, are treated similarly to ensure that they execute all intermediary jumps.
  
\noindent\textbf{5)}
Although the trampolines are prepared during compilation, they are randomized at run-time inside the enclave. The randomization control flow is that it does not reveal the randomization pattern. Randomizing the trampolines forces the attacker to shadow all possible locations in the enclave and thus, prevents the attacker from shadowing the trampoline branches and reliably tracking the program's execution.

\noindent\textbf{6)}
An attacker could repeatedly call same enclave functionality to gradually determine the randomization pattern.
To prevent this we perform re-randomization of trampolines.
This is done on all enclave entries, but is currently not integrated to our system.
For future work we envision supporting re-randomization using programmer annotations and compiler-based analysis (e.g., to locate repetitive secret dependent branches).

\section{Implementation details}
\label{sec:imp-details}
We have implemented an open-source prototype of our approach, based on LLVM 6.0 and implemented in the X86 target backend. The instrumentation is applied by systematically traversing all functions and modifying their branching instructions, as explained in Section~\ref{sec:proposed-scheme}.
Since the run-time randomization library itself cannot be randomized, it must be resistant to branch-shadowing attacks.
While implemented, we have not yet integrated the randomizer to our instrumentation.
For efficient and fine-grained randomization we do not preform in-place randomization, instead, we move trampoline entries between two trampoline areas.
\ifdefined\arxiv
Listing~\ref{lst:shuffle} shows an overview of our randomization algorithm.
\else
Listing~\ref{lst:shuffle} shows an overview of our randomization algorithm, and a detailed description is available in our extended technical report~\cite{techrep}.
\fi

\lstset{language=C} 
\begin{lstlisting}[float,floatplacement=t,label=lst:shuffle,caption=Randomization algorithm]
for (entry = 0 : jump_table_entries) {
    location = rand() % trampoline_size;
    if (fits(entry, location)) {
        mark_reserved(location, entry);
    } else {
        l = (l+1) % trampoline_size;
    } 
    move_entry(entry, location);
}
\end{lstlisting} 

We have also implemented an application for shadowing in-enclave execution in a controlled manner.
Our setup is similar to~\cite{SGX-Branch-Shadowing} i.e., our application 1) retrieves branch instruction addresses and sets up a corresponding shadow-jump, 2) executes the attacked enclave function and returns, 3) enables performance counters and executes the shadow-code, and 4) reads performance counters to infer in-enclave execution.
Our setup is such that it could be integrated into the SGXStep-framework.
We have replicated the shadowing techniques shown by~\cite{SGX-Branch-Shadowing} and also performed shadowing on return statements.

We minimize performance overhead by performing costly operation only when needed. 
In particular, we: a) provide code-annotation for limiting the obfuscation to only developer-determined sensitive parts, and b) randomize the trampoline code only when detecting multiple enclave (re)entries (i.e., after a given number of potential shadowing attempts).

\section{Evaluation}
\label{sec:evaluation}

\subsection{Security Analysis}

As specified in Requirements (Section~\ref{sec:threat-model}), we must prevent an attacker from inferring the secret-dependent control flow by \ref{reqObservable}) ensuring that observable branches do not leak information, and \ref{reqNonObservable}) preventing the attacker from probing other branches with a probability based on the security parameter $\mathtt{k}$.

To hide any data-dependant branches (\ref{reqObservable}), we replace all conditional branches with unconditional branches. 
We further setup the control flow so that each block in the static code section is executed in the same order and on each function call. 
One limitation is that we do not conceal the number of loop executions, because this can be unknown compile time.
While unimplemented, in some cases this could be avoided by unrolling loops.

The remaining branching instructions are exclusively in the trampolines, for which the locations are randomized to defend against shadowing (\ref{reqNonObservable}).
Without knowing the exact trampoline layout, the attacker is forced to guess or exhaustively probe all possible locations.
The probability of attack success ($P_{\mathit{attack}}$) is given by Equation~\ref{eq:attackProb}, where $\mathtt{G}$ is the number of guesses and $\mathtt{k}$ the number of possible trampoline locations.

\begin{equation}
    \label{eq:attackProb}
        P_{\mathit{attack}} = \frac{G}{k}
\end{equation}

The upper limit for $\mathtt{G}$ is the number of BTB entries, but in practice this is lowered by any intermediate code (e.g., system calls and attack setup) that pollutes the BTB. 
The security parameter $\mathtt{k}$ determines the trampoline randomization space.
Because X86 allows unaligned execution, a single 4KB range gives us up to 4091 potential trampoline locations (with a trampoline size ranging from 5 to 15 bytes).
With a randomization area of 8KB and 4096 BTB entries, the attacker's success probability of shadowing a single branch has an upper bound of 0.5. The probability of following the full control flow thus drops exponentially as the number of targeted branches increase.

\subsection{Performance Evaluation}
We measured the performance overhead of our system in terms of CPU-utilization, memory use, and code size. 
Our experiments were conducted on an SGX-enabled Intel Skylake Core i5-6500 CPU, running at 3.20~GHz, with 7,6~GiB of RAM.
The machine was running Ubuntu 16.04 with 64-bit Linux 4.4.0-96-generic kernel and the SGX SDK version 2.0.
 
In order to evaluate the performance of our approach, we used SGX-Nbench~\cite{SGX-nbench} which is adapted from Nbench-byte-2.2.3, to measure the performance of 10 different benchmarks executed within an enclave. 
In Table~\ref{experiment-result}, we present the performance overhead introduced by our approach.
All benchmarks were conducted with full instrumentation, but do not include randomization or dummy-instructions, both of which we still to be implemented. 
Although the randomization will introduce overhead, it need not be constantly repeated.
Instead it can be performed once on enclave creation and then later after a specified number of enclave re-entries.

\begin{table}[h]
\caption{CPU overhead introduced by our approach. Numbers are reported as iterations/second before and after instrumentation. The overhead introduced by randomization and dummy instructions are not included in this experiment.}{
\label{experiment-result}
\begin{tabular}{p{2.3cm}p{1.5cm}p{1.5cm}p{2cm}} 
\hline
\centering Benchmark & \centering Before\\(std. dev.) & \centering After\\(std. dev.)& \centering Performance loss\tabularnewline \hline \hline
Numeric sort &\centering 828.8 (0.79) &\centering 578.8 (0.21) &\centering 30\%  \tabularnewline
String sort &\centering 86.59 (0.09) &\centering 67.72 (0.21) &\centering 21\% \tabularnewline
Bitfield  &\centering  1.839e8 (1.34e5) &\centering  1.370e8 (3.27e5)&\centering 25\%  \tabularnewline 
Fp emulation     &\centering 87.70 (0.11) &\centering 42.73 (0.02) &\centering  51\%    \tabularnewline
Fourier          &\centering  1.789e5 (1.19e2) &\centering  1.500e5 (1.50e2) &\centering   16\% \tabularnewline
Assignment       &\centering 21.64 (0.03) &\centering  7.769 (0.01) &\centering 64\%\tabularnewline
Idea             &\centering 2667 (1.26) &\centering  2665 (1.84) &\centering   0.1\%     \tabularnewline
Huffman          &\centering 2354 (4.07) &\centering  860.5 (0.71) &\centering   63\%    \tabularnewline
Neural net       &\centering 35.16 (0.03) &\centering 25.57 (0.22) &\centering  27\%\tabularnewline
Lu decomposition &\centering 973.1 (1.45) &\centering  785.0 (1.41) &\centering   19\%   \tabularnewline
\hline
Average & & & \centering 17.17\% \tabularnewline
\hline
\end{tabular}}
\end{table}

\noindent\textbf{CPU overhead:} The increase in execution time is caused by the added trampoline jumps and the need to exhaustively execute all jump-blocks. 
Table~\ref{experiment-result} presents the execution time of various benchmarks in the enclave, before and after obfuscating them. 
The decrease in the number of iterations per second clearly shows a non-negligible performance degradation.
However, since we have obfuscated the entire program in these benchmarks, these represent the worst case performance overheads.
In real deployments, we would obfuscate only the parts of the code that depend on secret enclave data. The performance loss range can be grouped as very low (3-16\%), middle (25-32\%), and high (60\% and above). This variation depends on how complicated the function is in terms of size and number of branches. The Idea benchmark, for instance, has functions with many nested conditional branches all of which require corresponding jump-blocks to be added and executed.
 
\noindent\textbf{Memory overhead:} As expected, our instrumentation does not increase heap or stack use, but does increase the code size that must be loaded within the enclave.
For measuring the code size, we compared the size of the enclave object files before and after instrumentation. 
The code size of SGX-Nbench increased 1.6 times (from 39.6 kB to 64.6 kB) after instrumentation.

\section{Related work}
\label{sec:related-work}

There is a growing body of research on side channel attacks targeting Intel SGX and corresponding countermeasures. 
In addition to the branch-shadowing attacks~\cite{SGX-Branch-Shadowing, BranchScope} that we discussed in detail, there other side channel attack targeting SGX enclaves~\cite{Cache-Attack1, cache-Attack2, Controlled-Channel-Attacks, SGX-page-table-based-attacks}. In order to mitigate these side channel attacks, there have been various approaches proposed.
In the following, we discuss these countermeasures and how they differ from our work.

 Several research works have been presented to thwart controlled-channel (page-fault) attacks.
\textit{SGX-Shield}~\cite{SGX-Shield} randomizes the memory layout, similar to Address Space Layout Randomization (ASLR), in order to prevent control flow hijacking and hide the enclave memory layout.
This approach impedes run-time attacks that exploit memory errors or attacks that rely on a known memory layout (e.g., controlled-channel attacks). 
SGX-shield uses on-load randomization, allowing repeated branch-shadowing attacks to gradually reveal the randomization pattern.
Our approach solves this through run-time re-randomization. 
We further minimize the additional attack-surface by limiting the randomization to the trampolines, not the whole code section.

Shinde et al.~\cite{Defense-against-page-fault} propose an approach that masks page-fault patterns by making the program's memory access pattern deterministic. 
More precisely, they alter the program such that it accesses all its data and code pages in the same sequence, regardless of the input. 
This makes the enclave application demonstrate the same page-fault pattern for any secret input variables. 
T-SGX~\cite{T-SGX} is another solution proposed for mitigating page-fault attacks. 
It leverages Intel Transactional Synchronization Extensions (TSX) to suppress encountered page-faults without invoking the underlying OS. T-SGX does not mitigate the branch-shadowing attack \cite{SGX-Branch-Shadowing}, but it could be combined with our approach to address both branch-shadowing and page-fault attacks.

\textit{DR.SGX}~\cite{DR-SGX} is presented to defend against cache side-channel attacks. 
It permutes data locations, and continuously re-randomizes enclave data in order to hamper correlation of memory accesses. 
This approach prevents leakages resulting from secret-dependant data accesses. 
Chandra et. al~\cite{Data-Analytics-Randomization} propose a defense against side-channel attacks on SGX and for protecting data privacy. 
To do this, dummy data instances are injected into the user-given data instances in order to add noise to memory access traces. 
Moreover, they randomize/shuffle the dummy data with the user data to reduce the chance of extracting sensitive information from side-channels.
Both approaches are similar to our approach in that they employ randomization; however, the target of randomization is data and they do not defend against the branch-shadowing attack.

CCFIR (Compact Control Flow Integrity and Randomization)~\cite{CFI-Randomization} is a new method proposed to impede control-flow hijacking attacks (e.g., returninto-libc and ROP). 
CCFIR controls the indirect control transfers and limits the possible jump location to a whitelist in a Springboard. 
Randomizing the order of the stubs in the Springboard adds an extra layer of protection and frustrates guessing of the function pointers and return addresses. 
The environment for which CCFIR is proposed differs from ours.

Obfuscation techniques were previously used to thwart leakage via side-channel attacks. 
Oblivious RAM (ORAM)~\cite{ORAM} continuously shuffles and re-encrypts the data when accessed in memory, disk or from server, in order to conceal the program's memory access patterns. 
It requires that the state is stored and updated client-side. 
This makes it difficult to use ORAM for cache protection purposes, as it is challenging to store the internal state of ORAM securely without hardware support, given the small size of cache lines. 
Moreover, this approach suffers from considerable performance overhead.

None of the above countermeasures focus on mitigating branch-shadowing attacks, and additionally, Lee et. al~\cite{SGX-Branch-Shadowing} have demonstrated that their designed branch-shadowing attack is capable of breaking the security constructs of SGX-Shield, T-SGX, and ORAM.

\section{Conclusion and Future work}
\label{sec:conclusion}
 
We propose a software-based mitigation scheme to defend against branch-shadowing attacks, even in the presence of attackers with the ability to single-step through SGX enclaves.
Our approach combines compile-time control flow obfuscation with run-time code randomization to prevent the enclave program from leaking secret-dependant control flow. 
To evaluate the performance of our approach, we measured the run-time overhead of ten benchmark programs of SGX-Nbench.
Although we evaluated the worst-case scenario (whole program instrumentation), our results show that, on average, our approach results in a 1.6 times code size increase and less than 18\% performance loss.

As the future work, we will implement the randomizing component, and optimize our obfuscator to reduce overhead. 
In addition, we will improve our approach by integrating existing defences, in order to address other side-channel attacks.

\balance

\bibliographystyle{ACM-Reference-Format}
\bibliography{ccs-sample}

\end{document}